\documentclass[aps,prl,twocolumn,showpacs,superscriptaddress]{revtex4-1}
\usepackage{graphicx}
\usepackage{dcolumn}
\usepackage{bm}
\usepackage{amssymb}
\input epsf

\begin{document}

\title{Mimicking the LCDM model with Stealths}

\author{Cuauhtemoc Campuzano}
\email{ccampuzano[at]uv.mx}

\affiliation{ Facultad de F\'isica, Universidad Veracruzana, 91000, Xalapa
Veracruz, Mexico}

\author{V\'{\i}ctor H. C\'ardenas}
\email{victor.cardenas[at]uv.cl}

\affiliation{Instituto de F\'{\i}sica y Astronom\'{\i}a, Facultad
de Ciencias, Universidad de Valpara\'{\i}so, Av. Gran Breta\~na 1111,
Valpara\'{\i}so, Chile}

\author{Ram\'on Herrera}
\email{rherrera[at]ucv.cl}
\affiliation{Instituto de F\'{\i}sica, Pontificia Universidad Cat\'{o}lica de
Valpara\'{\i}so, Avenida Brasil 2950, Casilla 4059,
Valpara\'{\i}so, Chile.}

\begin{abstract}
We present a new cosmological model that mimics the Lambda Cold Dark
Matter by using a stealth field. This kind of field is
characterized as not coupling directly to
gravity; however, it is connected to the underlying matter content of the
universe model. As is known, stealth fields do not
back-react on the space-time; however, their mimicry skills show how
this field and its self-interaction potential determines the cosmic
evolution. We show the study of the simplest model that can be developed with the
stealth field.
   \end{abstract}

\maketitle

\textit{I. Introduction}. Precise astronomical measurements of the
universe indicate that nearly $25 \%$ of its content is in the
form of dark matter (DM), the key ingredient necessary to explain large scale structure formation, and $70\%$ dark energy (DE), the
 unknown component driving the recent cosmic acceleration.

In this context, the best model to describe almost all the
observational data is a mixture of elements from the standard
cosmological model plus a cosmological constant, the so-called
``concordance'' $\Lambda$CDM model. Although successful in fitting
the observational data, from a theoretical point of view the model
seems too arbitrary. First, there is no clue about where this
cosmological constant came from, and with it, why its value is so
close to the critical energy density, and second, why we live in a
very special epoch where the contributions from DM and DE
are of the same order of magnitude, the well-known ``cosmic
coincidence problem'' (CCP).

Physicists have proposed different ways to overcome this
dilemma. The first was to adopt a dynamical cosmological constant,
trying to adjust the dynamics of it to alleviate the CCP. This is
the idea behind quintessence
\cite{quinta1,quinta2,quinta3,quinta4,quinta5,quinta6}, where a
scalar field is responsible for driving the current cosmic
acceleration. The second was to modify the left-hand side of Einstein's
equations, trying to explain the presence of a cosmological constant
as a non-standard geometric effect \cite{Tsujikawa2010,
Capozziello2011, Starkman2011}. The third was to violate the
Copernican Principle, i.e., by assuming we live in an inhomogeneous
universe \cite{ltb1,ltb2,ltb3,ltb4}. Although some successes have
been obtained in each one of these alternatives scenarios, there is
so far no clear evidence of a preference compared to the
$\Lambda$CDM model.

Since Einstein's field equations link the geometric properties of
the universe with its total content, there is a well-known
degeneracy between these two components; DM and DE.
This is in fact one good reason to consider unified dark models. Of
course the \textit{simplicity} of considering a single component acting
as both DM and DE is also a good reason.

In this Letter, we show that a new class of scalar field model that
exhibits a non-trivial response to geometry, dubbed
\textit{stealth}, which serves as a unified
model of the dark sector.

The idea of considering unified scalar field models (see \cite{udm}
for a review) to describe DM and DE emerges
as a natural way to alleviate the so-called ``coincidence problem'',
namely, to explain why the energy densities of these two dark
components are of the same order of magnitude today. Models of this
type have been proposed in the past, assuming the stress energy
tensor of the scalar field back-react to the geometry according to
Einstein's equations. Among them we can mention the model of
\cite{sahni2000}; where a potential $V(\phi)=V_0 (\cosh \lambda \phi
- 1)^p$ is considered, the Chaplygin gas \cite{chaplin}, the
generalized Chaplygin gas model \cite{chaplinG}, and models with a
non-canonical kinetic term called k-essence models \cite{kessence}.
%

%
On the other hand, it is well-known that in The General theory of Relativity gravity is understood as a manifestation of the curvature of space-time and the latter is
caused by the presence of matter. This fundamental principle
is codified in the equations proposed by Einstein.
So the slightest presence of matter on the right hand
side of Einstein's equations is sufficient to alter the geometry
of space-time. However, the stealth is a kind of matter that remains present in the space-time without altering or changing its geometry. The stealth appears only for a scalar field non-minimally coupled to gravity,  
and its origins date back to the improved energy-momentum tensor considered first in \cite{Callan70}, where the authors showed the possibility of this new
 tensor becoming the source of the gravitational field; meanwhile, the dynamic of the scalar field is dictated by the Klein-Gordon equation.


%

The original proposal of the stealth was
reported for a three-dimensional BTZ black hole in
\cite{AyonBeato:2004ig}, in higher dimensions in Minkowski space-time
\cite{AyonBeato:2005tu} and (anti-)de Sitter [(A)dS] space
\cite{eloy2014}. Also in Lifshitz space-time in \cite{mokthar}, for a four dimensional
black hole \cite{mokthar2}, for an AdS black hole in Lovelock gravity
\cite{mokthar3}, in Einstein-Gauss-Bonnet gravity for topological
black hole \cite{mokthar4}, for a rotating AdS black hole in new massive
gravity \cite{mokthar5}, and finally for a BTZ rotating black hole present two solutions in \cite{mokthar6}. Lately, as was shown in \cite{ayon2015}, there are stealth fields during the cosmological evolution and some cosmological solutions are given in order to have a LCDM cosmology, in particular those with polynomials and power-law evolution are analyzed. Also, the general solutions for de Sitter cosmologies and inhomogeneous stealths have been studied, concluding that only for de Sitter backgrounds allow a full dependence on the space-time coordinates.
In this letter, we examine the case of a cosmological model coming
from a non-minimal coupling with a stealth scalar field as a
unified component describing both DM and DE, and thus mimicking the $\Lambda$CDM model.

\textit{II. Stealths as a Unified dark model}.

In the present work we study a cosmological model coming from a non-minimal
coupling with a stealth scalar field, described by the action:
\begin{eqnarray}\label{action}
S=\int d^4x \sqrt{-g}\left[ \frac{R}{2\kappa} + L_m
- \frac{1}{2}\zeta R \phi^2
-\frac{1}{2}\partial_{\mu} \phi \partial^{\mu} \phi
- V(\phi)\right].\nonumber\\
\end{eqnarray}
Here, $L_m$ is the Lagrangian matter. Clearly, for $\zeta=0$ the
scalar field stress tensor reduces to the usual case of a minimally
coupled field. By varying the action (\ref{action}), the field
equations are written as
\begin{eqnarray}\label{feq}
 G_{\mu\nu}-\kappa T^{(m)}_{\mu\nu}= \kappa T^{(S)}_{\mu\nu},
\end{eqnarray}
where $T^{(m)}_{\mu\nu}$ is the stress energy tensor of matter, and
$T^{(S)}_{\mu\nu}$ is the stress energy tensor of the stealth field
$\phi$ given by
\begin{eqnarray}\label{stealtuv}
T^{S}_{\mu\nu}&=&\nabla_\mu \phi \nabla_\nu \phi -\left(V(\phi)+
   \frac{1}{2}\nabla_\alpha \phi \nabla^\alpha \phi\right) g_{\mu\nu}
   \nonumber\\&+&
  \zeta (G_{\mu\nu}\phi^2-\nabla_\mu \nabla_\nu \phi^2
  +g_{\mu\nu} \nabla^\alpha \nabla_\alpha \phi^2 ).
\end{eqnarray}
It is worth noting that for $\zeta \neq 0$, the variation on
$g_{\mu\nu}$ produces the stealth stress tensor $T^{(S)}_{\mu\nu}$ to 
get a contribution from the Einstein tensor.

The stealth configuration emerges once we set both sides of Eq.(\ref{feq})
to zero: the left-hand side is the Einstein's equation for a universe
with a matter content described by $T^{(m)}_{\mu\nu}$, and the right-hand
side is the stealth equations $T^{(S)}_{\mu\nu}=0$. Once a solution to
the right side is found, the stealth obeys the dynamics dictated by the
space-time and at the same time it is invisible to it.

While on the one hand the existence of gravitational stealth is a fact, and its feature of not having back-reaction on the gravitational field is of interest, the gravitational field equations say very little about their interaction with matter.
%
There are a few works on this topic; some remarkable results in that direction are given in \cite{sokolowski}, where the interaction between ordinary matter
and stealth is shown and \cite{mokthar7}, where a relation with the axionic field is shown. Furthermore, it is possible to show the ability of the stealth fields to mimic any kind of matter, which is another surprising characteristic of these fields \cite{temoc}.
%

%

%

As was demonstrated in \cite{Ayon-Beato:2013bsa}, there is a
stealth solution in the context of a Friedmann-Lemaitre-Robertson-Walker (FLRW) space-time. Now we obtain our cosmological model by coupling
\begin{equation}\label{frw}
  ds^2 = -dt^2 + a(t)^2 \left[\frac{dr^2}{1-k r^2}
  +r^2 d\Omega^2 \right],
\end{equation}
to a perfect fluid with zero pressure, i.e., the dust case, plus a cosmological constant, and when the scalar field depend only on time.

On the left-hand side of (\ref{feq}) we use the stress-energy tensor for a
perfect fluid as the DM contribution, leading to the usual
$\Lambda$CDM model, where the DM density $\rho$ and the
cosmological constant $\Lambda$ determine the cosmic evolution $a(t)$.
At the same time, from the right-hand side of (\ref{feq}), the stealth field $\phi$ and its self interacting potential $V(\phi)$ determines completely the cosmic evolution
$a(t)$. As a consequence of this -- the evolution must be the {\it same} as that of the $\Lambda$CDM model -- the stealth here works as a unifying field simultaneously describing
the effects of the action of both DM and DE. We get:
\begin{equation}\label{eq1}
3\left(\frac{\dot{a}}{a}\right)^2 + 3\frac{k}{a^2} =
-6\frac{\dot{\phi}}{\phi}\frac{\dot{a}}{a}
  -\frac{1}{2\zeta}
  \left(\frac{\dot{\phi}}{\phi}\right)^2-\frac{V}{\zeta\phi^2},
\end{equation}
and
\begin{equation}\label{eq2}
\frac{2V}{3\zeta\phi^2}+
\frac{6\zeta-2}{3\zeta}\left(\frac{\dot{\phi}}{\phi}\right)^2
+2\frac{\dot{\phi}}{\phi}\frac{\dot{a}}{a}+
2\frac{\ddot{a}}{a}+2\frac{\ddot{\phi}}{\phi}=0.
\end{equation}

\textit{ III. Specific Stealth realizations}

As is well-known, the $\Lambda$CDM model is so far the best fit
model for a large set of astronomical observations, such as, type Ia
supernovae (SNIa), baryon acoustic oscillations (BAO), cosmic
microwave background radiation (CMBR), growth of structure, etc.
\citep{pero}. In this setup the cosmological constant $\Lambda$
drives the current accelerated expansion of the universe, detected
for the first time using type Ia supernovae \citep{snia1},
\citep{snia2}.

As we mentioned in the introduction, although the stealth field
does not back-react to the geometry, the existence of a non-zero
coupling $\zeta$ enables the stealth to appear dynamically coupled
to the matter content. In this section we characterize the stealth
field associated with this cosmological model.

In order to give a complete description of the model we are
presenting, we display the features of the $\Lambda$CDM model.
The Friedmann equation is
\begin{equation}
  H^2+\frac{k}{a^2} = \frac{\kappa}{3} \rho + \frac{\Lambda}{3},
\end{equation}
and the stress-energy conservation equation implies
\begin{eqnarray}
  \dot{\rho}+3H\rho=0,\label{8}
\end{eqnarray}
where we have assumed explicitly an equation of state $p=0$ for the
matter content (cold dark matter). From equation (\ref{8}) we obtain the
energy density, which evolves as $\rho = \rho_0 a^{-3}$. This simple model
fits several observational probes quite well. The best fit parameters
so far, assuming a curved FRW metric, are those from the Planck
Collaboration \cite{planck}: $\Omega_{\Lambda}=0.685 \pm 0.018$,
$\Omega_m=0.315\pm 0.018$, and $H_0=67.3 \pm 1.2$, where
$\Omega_{\Lambda}=\Lambda/(3H_0^2)$, $\Omega_k=-k/H_0^2$, and
$\Omega_m=\kappa\rho_0/(3H_0^2)$.

On the other hand, from the vanishing of the stealth stress-energy
tensor, Eqs.(\ref{eq1}, \ref{eq2}), we can read the equivalence
relations between the set $[\rho, \Lambda]$ for the $\Lambda$CDM
model, and the set $[\phi, V(\phi)]$ for the stealths.
It is easy to show that an equivalence can be met by proposing the
following relation:
\begin{equation}\label{potfix}
-\frac{V}{\zeta\phi^2}=\Lambda.
\end{equation}
This means that the self interacting potential is related only
algebraically to the cosmological constant.
Using this relation and after some manipulations, the equivalence is
complete after we impose
\begin{equation}\label{2ndrel}
- \frac{\dot{\phi}}{\phi} \frac{d}{dt}\ln [\phi^{1/2\zeta} a^6] =
\kappa \rho,
\end{equation}
as well as
\begin{equation}\label{key}
\frac{\dot{\phi}}{\phi} = - \frac{c}{\phi^{\beta}a^2 },
\end{equation}
where $\beta = (8\zeta-1)/4\zeta$, and $c$ is an integration
constant.

What we have obtained here is a stealth field $\phi$ -- given by the
solution of (\ref{key}) -- with self-interaction (\ref{potfix}) that
(by construction) {\it generates} an evolution -- $a(t)$ -- totally
indistinguishable from that obtained from the $\Lambda$CDM model.

The equivalence enables us to use cosmological observations to fix
the values (and the uncertainties) of the model parameters. Using
(\ref{key}) in the equation for $\rho$ we get for the Hubble function:
\begin{equation}\label{heq}
H = \frac{\kappa \rho_0}{6a}\frac{\phi^{\beta}}{c} +
\frac{1}{12\zeta a^2}\frac{c}{\phi^{\beta}}.
\end{equation}
Before testing the model, we have to write it in terms of the redshift
$z$. Recalling that $a=(1+z)^{-1}$ the relation (\ref{key}) can be
written as


\begin{equation}\label{phiz}
E(z) \varphi^{\delta}\varphi ' = \frac{(1+z)\Omega_{m}}{\Xi},
\end{equation}
where $E(z)=H(z)/H_0$, $\varphi=\phi/\phi_0$, $\delta=\beta -1$ and
$\Xi=1 \pm \sqrt{1-\Omega_m/6\zeta}$. On the other hand,
Eq.(\ref{heq}) can be written as
\begin{equation}\label{heq2}
E(z) = \frac{(1+z)}{2}\varphi^{\beta} \Xi + \frac{(1+z)^2}{12\zeta
}\frac{\Omega_m}{\varphi^{\beta} \Xi}.
\end{equation}
The free parameters to constrain are clearly $\zeta$ and
$\Omega_m$. There is no way to constrain $H_0$ based on sets
(\ref{phiz}) and (\ref{heq2}). However, if we test the model using
$H(z)$ measurements, we can get a number for $H_0$ just by
minimizing the residuals of
\begin{equation}\label{resi}
\left[H^{obs}(z_i) - H_0 E(z_i | \zeta,\Omega_m)\right].
\end{equation}
In practice we solve differential equation
(\ref{phiz}) numerically with the initial condition $\varphi(z=0)=1$, and by
making use of (\ref{heq2}) to get $E(z)$. Then, we compute the
residuals. In what follows, we use observational measurements of
$H(z)$ extracted from \cite{hdz} -- consisting of 30 data points --
to constrain the free parameters in the model.

In addition to the parameters $\zeta$, $\Omega_m$ and $H_0$, we must also
consider the parameters associated with the stealth potential
$V(\phi)$. Given our choice of (\ref{potfix}), the potential can
be described by just one parameter that is fixed by relation
(\ref{potfix}). In fact, by writing $V=V_0 \phi^2$, from
(\ref{potfix}) we find that $V_0=-3H_0^2\zeta \Omega_{\Lambda}$. In
this way, it is not necessary to fit it along with the other three,
because it depends on the best fit value of $\zeta$, $H_0$ and
should be consistent with the known value of $\Omega_{\Lambda}$.

\begin{figure}
\centering
\begin{tabular}{c}
\includegraphics[width=6cm]{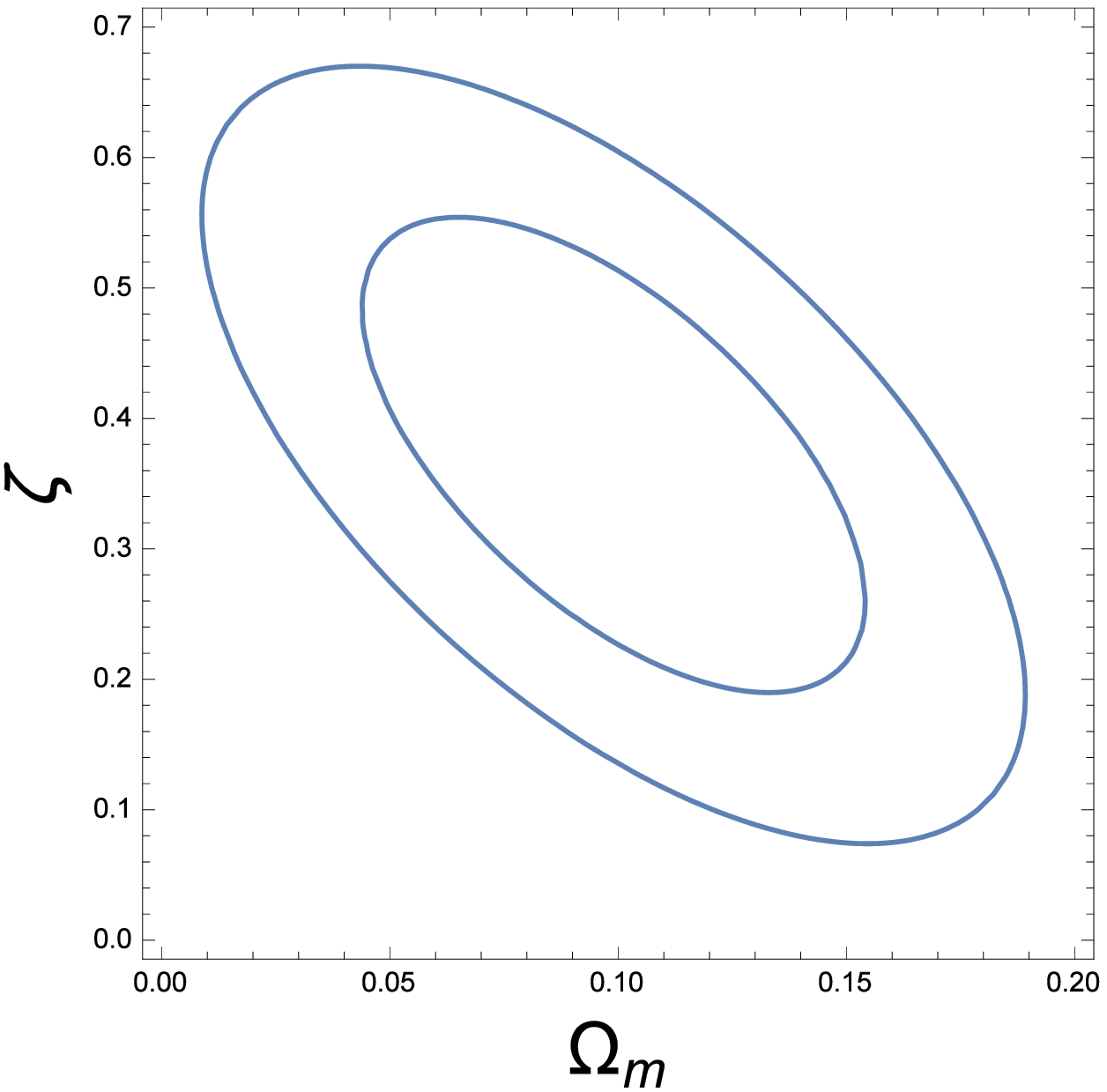} \\
\includegraphics[width=6cm]{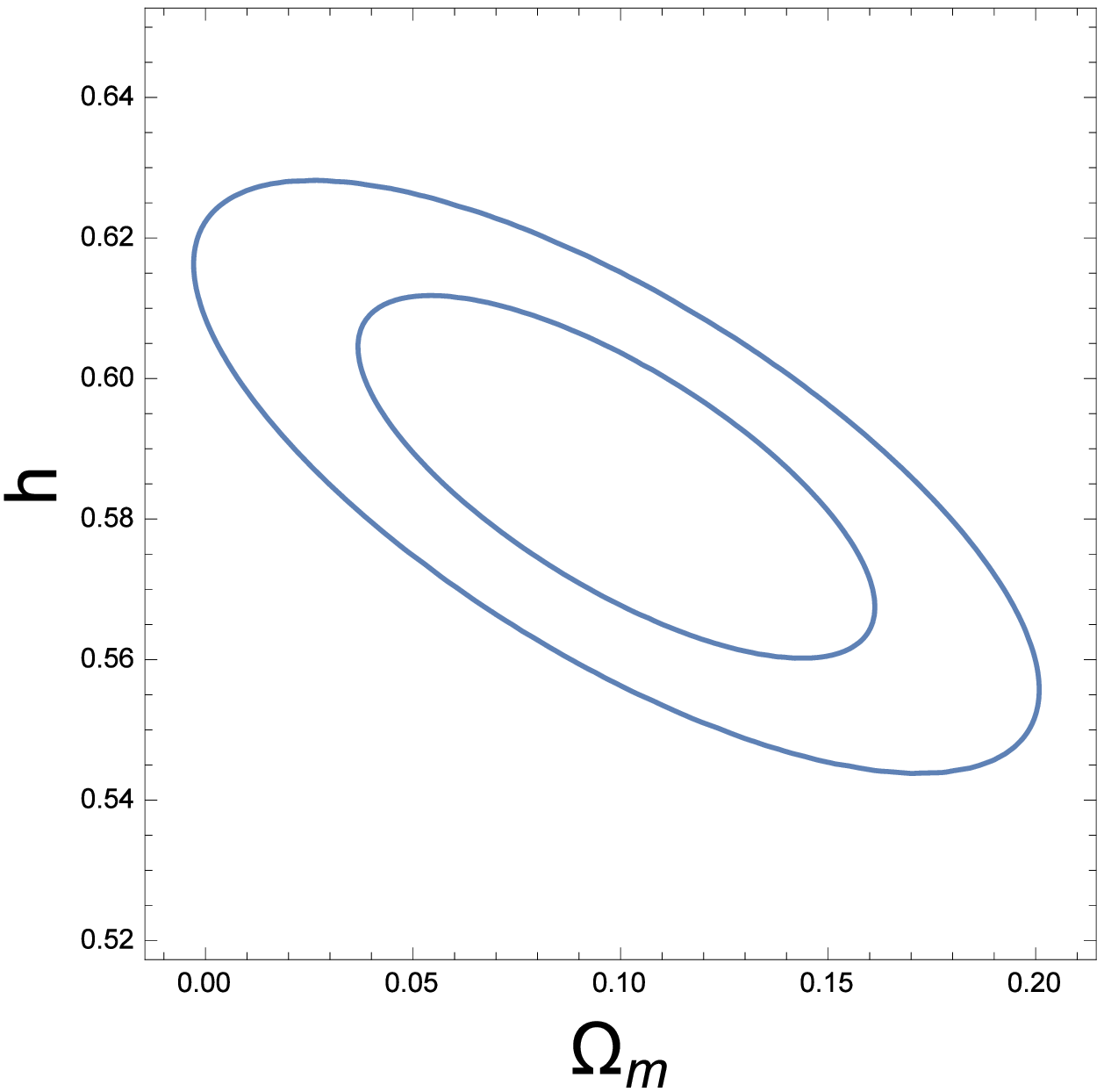} \\
\includegraphics[width=6cm]{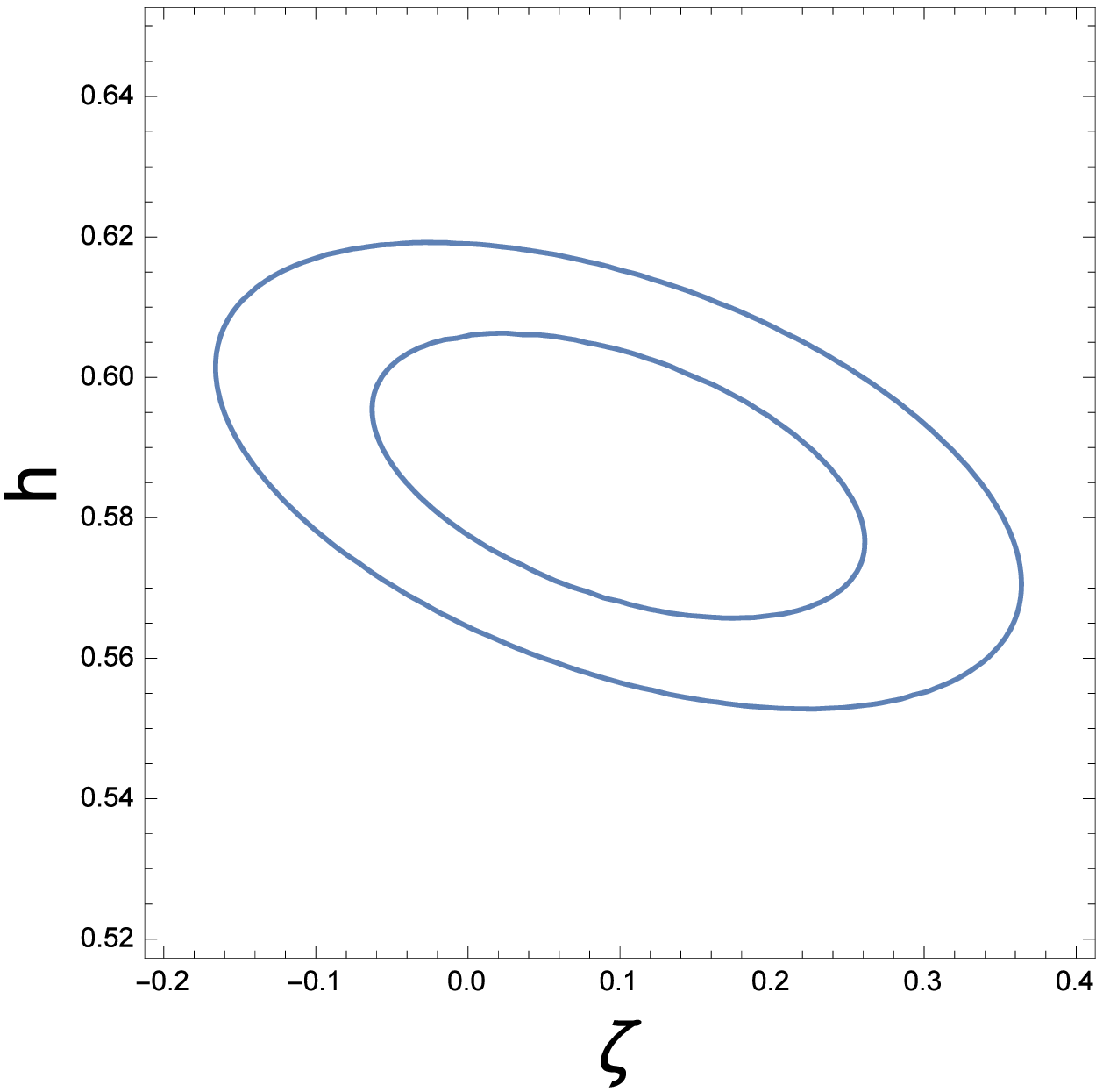}
\end{tabular}

\caption{Here we display the confidence level contours, at $1
\sigma$ and $2 \sigma$, for the parameters of the
model.}\label{pdfz}
\end{figure}

\begin{figure}
 \begin{center}
    \includegraphics[width=8cm]{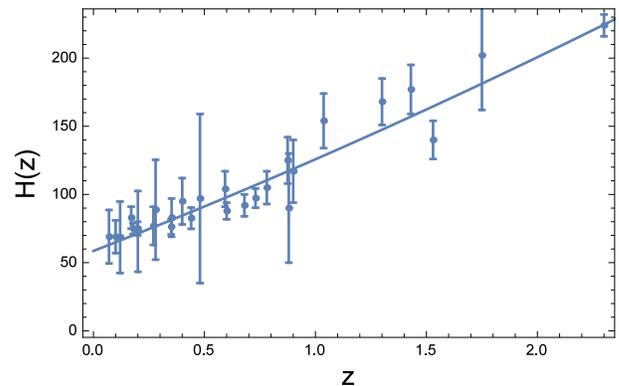}
  \end{center}
\caption{We display the theoretical curve together with the data points
for the $H(z)$ measurements obtained from \cite{hdz}. }\label{fitH}
\end{figure}

Along these lines, it is clear that our Stealth have more free
parameters than the original $\Lambda$CDM model (the former has four
and the latter three). However, as we just mentioned, the only free
parameters that we can fix using (\ref{phiz}) and (\ref{heq2}) are
$\zeta$, $\Omega_m$ and $H_0$. After the fit (assuming the plus sign in $\Xi$), we get $h=0.59 \pm
0.02$, $\Omega_m = 0.10 \pm 0.05$ and $ \zeta = 0.10 \pm 0.12$. In
Fig.(\ref{pdfz}) we show the $1 \sigma$ and $2 \sigma$ C.L. among
the free parameters. In Fig. (\ref{fitH}) we display the data points
used to constrain the model along with the best theoretical curve.

The reduced Hubble parameter ($h=H_0/100$) is the most sensitive
parameter in the fit. As we mentioned in the last paragraph, this
parameter essentially controls the amplitude of the theoretical
curve displayed in Fig.(\ref{fitH}). By contrast parameters $\Omega_m$ and
$\zeta$ are not very sensitive to changes, so it was very difficult to find a best fit set based on the $H(z)$ data. A close study of the system of Eqs.(\ref{phiz}) and (\ref{heq2}) enables us to understand this behavior. In fact, the use of observational data for each value of $E(z)$ -- instead of using an analytical expression for it -- renders these two parameters highly correlated.

\textit{IV. Conclusions}

In this paper we have shown an example of how the Stealth can operate during the cosmological evolution describing the $\Lambda$CDM model. This example make use of an explicit quadratic potential for the Stealth, and opens the possibility of extends this finding using other forms of $V(\phi)$. We have also re-write the Stealth equations in a way to find explicitly the equivalence between the $\Lambda$CDM content -- non-relativistic matter and a cosmological constant -- and the Stealth field and its potential. It was through this philosophy -- working with the Stealth equivalent of the $\Lambda$CDM -- that we have found the example we studied.

To put our model to the test, considering the special features of the current cosmological model, we opt not to use an ansatz for the scale factor $a(t)$; instead, we make use of observational data directly to constrain the free parameters of the stealth.
What we have shown here is the {\it observationally} induced stealth that best
describes the $\Lambda$CDM model. Here, the stealth field with
its self-interacting potential enables us to describe both the dark
matter and the cosmological constant contributions, thus being a unified scalar field model. The best fit curve -- i.e., the function $E(z)$ extracted from (\ref{heq2}) -- together with  the data is shown in Fig. (\ref{fitH}), showing the capacity of the stealth to describe the observational data directly.

Although by construction the stealth mimics the evolution of the
$\Lambda$CDM model, we have performed a direct test of the model
against observational data, and we have found that in this case the best fit modifies the value for $\Omega_m$ (instead of the typical $\simeq 0.27$, by our value $\simeq 0.1$) at the expense of fixing an extra parameter, $\zeta$, which is absent in the $\Lambda$CDM model.

%
%
%
Finally, we would like to emphasize the importance of understanding the potential role the stealth may play in cosmic evolution. As we have shown, although stealth does not back-react to the space-time, it can describe both dark contributions at once, accounting for almost 95\% of the matter content of the universe. There is no doubt that we must continue to explore the consequences of the stealth in the recent cosmic evolution as well as in the early ages of the universe.



\section*{Acknowledgments}

This work is dedicated to the memory of Professor Sergio del Campo. CC
acknowledges partial support by CONACyT Grant CB-2012-177519-F. CC
also acknowledges CONACyT Grant I0010-2014-02 Estancias
Internacionales-233618-C. This work was partially supported by SNI
(M\'exico). VHC acknowledges partial support by grant DIUV 50/2013.
One of the authors (VHC) would like to thank the warm hospitality of the
Facultad de F\'isica at the Universidad Veracruzana in Xalapa, where
part of this work was carried out. RH was supported by the Comisi\'on
Nacional de Ciencias y Tecnolog\'ia of Chile through FONDECYT Grant
N$^{0}$ 1130628 and DI-PUCV N$^{0}$ 123.724.


\section*{References}
\begin{thebibliography}{9}

\bibitem{quinta1}
C. Wetterich, Nucl. Phys. B 302, 668, 1988

\bibitem{quinta2}
B. Ratra  and P.J.E. Peebles, Phys. Rev. D 37, 3406, 1988

\bibitem{quinta3}
J.A. Frieman , C.T. Hill , A. Stebbins  and I. Waga, Phys. Rev.
Lett. 75, 2077, 1995

\bibitem{quinta4}
M.S. Turner  and M. White, Phys. Rev. D 56, R4439, 1997

\bibitem{quinta5}
R.R. Caldwell , R. Dave  and P.J. Steinhardt, Phys. Rev. Lett. 76,
1582, 1998

\bibitem{quinta6}
P.J. Steinhardt, L. Wang , and I. Zlatev, Phys. Rev. D 59, 123504,
1999

\bibitem{Tsujikawa2010}
S.Tsujikawa, Lect.\ Notes Phys, 800, 99,  2010

\bibitem{Capozziello2011}
S. Capozziello and M. De Laurentis, Phys.\ Rept., 509, 167, 2011

\bibitem{Starkman2011}
G.~D. {Starkman}, Phil.\ Trans.\ Roy.\ Soc. Lond.\ A, 369, 5018,
2011

\bibitem{ltb1}
M. P. Dabrowski, Astrophys. J. 447, 43, (1995); M. P. Dabrowski and
M. A. Hendry, Astrophys. J. 498, 67, (1998).

\bibitem{ltb2}
J. F. Pascual-Sanchez, Mod. Phys. Lett. A14, 1539 (1999).

\bibitem{ltb3}
M.-N. Celerier, Astron. and Astrophys. 353, 63 (2000).

\bibitem{ltb4}
K. Tomita, Astrophys. J. 529, 382011 (2000); MNRAS 326, 287 (2001)

\bibitem{udm}
D. Bertacca, N. Bartolo, \& S. Matarrese, Advances in Astronomy,
2010, 904379 (2010).

\bibitem{sahni2000}
V.~Sahni and L.-M. Wang,
Phys. Rev. {\bf D62} 103517, (2000).

\bibitem{chaplin}
A.~Y. Kamenshchik, U.~Moschella, and V.~Pasquier,
Phys. Lett. {\bf B511}, 265--268,(2001).

\bibitem{chaplinG}
M.~C. Bento, O.~Bertolami, and A.~A. Sen,
Phys. Rev. {\bf D66}, 043507, (2002).

\bibitem{kessence}
R.~J. Scherrer, 
Phys. Rev. Lett. {\bf 93}, 011301, (2004).

\bibitem{Callan70}
C.G. Jr. Callan, S. Coleman, \& R. Jackiw, R., Annals of Physics,
59, 42 (1970).


\bibitem{AyonBeato:2004ig}
  E.~Ay\'on-Beato, C.~Mart\'{\i}nez and J.~Zanelli,
  Gen.\ Rel.\ Grav.\  {\bf 38}, 145 (2006)
  [hep-th/0403228].

\bibitem{AyonBeato:2005tu}
  E.~Ay\'on-Beato, C.~Mart\'{\i}nez, R.~Troncoso and J.~Zanelli,
  Phys.\ Rev.\ D {\bf 71}, 104037 (2005)
  [hep-th/0505086].

\bibitem{Ayon-Beato:2013bsa}
  E.~Ay\'on-Beato, A.~A.~Garc\'{\i}a, P.~I.~Ram\'{\i}rez-Baca
  and C.~A.~Terrero-Escalante,
  Phys.\ Rev.\ D {\bf 88}, no. 6, 063523 (2013)
  [arXiv:1307.6534 [gr-qc]].

\bibitem{eloy2014}
E. Ay\'onÂ–-Beato, C. Mart\'{\i}nez, R. Troncoso, and J. Zanelli, Â“Stealths
overflying (A)dS,Â” in preparation.

\bibitem{mokthar}
E. Ay\'on-Beato, M. Hassaine, M. M. Ju\'arez-Aubry.  [arXiv:1506.03545 [gr-qc]]

\bibitem{mokthar2}
M. M. Caldarelli, C. Charmousis, M. Hassaine.   JHEP 1310, 015 (2013).

\bibitem{mokthar3}
M. Bravo-Gaete, M. Hassaine. JHEP 1311, 177 (2013).

\bibitem{mokthar4}
M. Bravo-Gaete, M. Hassaine.
Phys.Rev. D88, 104011 (2013).

\bibitem{mokthar5}
M. Hassaine. Phys.Rev. D89, 4, 044009  (2014)

\bibitem{mokthar6}
M. Bravo-Gaete, M. Hassaine. Phys.Rev. D90 2, 024008 (2014).

\bibitem{ayon2015}
E. Ay\'on-Beato, P. I. Ram\'{\i}rez-Baca and C. A. Terrero-Escalante. (2015)
[arXiv:1512.09375[gr-qc]]. 

\bibitem{sokolowski}
 Sokolowski, L. M., Acta Phys.Polon. B35, 587, (2004).

\bibitem{mokthar7}
M. Bravo-Gaete, M. Hassaine. JHEP 10, 015 (2015).

\bibitem{temoc}
A. Alvarez, C. Campuzano, V. Cardenas, R. Herrera and E. Rojas
in preparation

\bibitem{banerj}
N. Banerjee, R. K. Jain and D. P. Jatkar, Gen. Rel. Grav. 40, 93
(2008) [arXiv:hep-th/0610109].

\bibitem{maeda}
H. Maeda and K.I. Maeda, Phys. Rev. D 86, 124045 (2012)
[arXiv:1208.5777 [gr-qc]].

\bibitem{ref5}
Ay\'on-Beato, Eloy et al. Phys.Rev. D88, 6, 063523 (2013)
arXiv:1307.6534

\bibitem{pero}
L. Perivolaropoulos, Journal of Physics: Conference Series, Volume
222, Issue 1, id. 012024 (2010).

\bibitem{snia2}
S. Perlmutter, et al., Astrophys. J. 517, 565 (1999)

\bibitem{snia1}
A.G. Riess, et al., Astron. J. 116, 1009 (1998).

\bibitem{planck}
P. Ade, et al. [Planck Collaboration] Astron. and Astrophys. (2014).

\bibitem{hdz}
  M.~Moresco {\it et al.},
  JCAP {\bf 1605}, no. 05, 014 (2016)
  doi:10.1088/1475-7516/2016/05/014
  [arXiv:1601.01701 [astro-ph.CO]].

\end{thebibliography}
\end{document}